\documentclass[a4paper,twoside,titlepage,twocolumn,fleqn,final,prb,aps]{revtex4}

 \usepackage{switch}  
 \usepackage[latin1]{inputenc}
 \usepackage{amsmath}
 \usepackage{amssymb}

\renewcommand{\S}{\mbox{\boldmath$S$}}
\newcommand{\R}{\mbox{\boldmath$R$}}
\newcommand{\rr}{\mbox{\boldmath$r$}}
\newcommand{\B}{\mbox{\boldmath$B$}}
\newcommand{\M}{\mbox{\boldmath$M$}}
\newcommand{\I}{\mbox{\boldmath$I$}}
\newcommand{\abs}[1]{\left| #1\right|} 
\newcommand{\VEV}[1]{\langle #1\rangle}
\renewcommand{\H}{\mbox{\boldmath$H$}}

\begin{document}

\title{Staggered magnetization, critical behavior and weak ferromagnetic properties of LaMnO$_3$ by muon spin rotation.} 

\author{M. Cestelli Guidi, G. Allodi, \href{http://www.fis.unipr.it/~derenzi/derenzi.html}{R. De Renzi}, G. Guidi}
\address{Dipartimento di Fisica e 
Istituto Nazionale di Fisica della Materia,\\ 
Universit\`a di Parma, I-43100 Parma, Italy }
\author{M. Hennion}
\address{Laboratoire L. Brillouin, CEA-CNRS, CE Saclay, F-91191 Gif sur Yvette , France} 
\author{L. Pinsard}
\address{Laboratoire de Chimie des Solides, Universit\'e Paris-Sud, F-91405, France}
\author{A. Amato}
\address{Lab. for Muon-Spin Spectroscopy, Paul Scherrer Institut, CH-5232 Villigen PSI,  Switzerland}

\date{\today}
\begin{abstract}
We present a study of a microtwinned single crystal of LaMnO$_3$ by means of implanted muons. Two muon stopping sites are identified from the symmetry of the internal field in the ordered phase. The temperature dependence of these fields yields the behavior of the staggered magnetization from which a static critical exponent ($\beta=0.36(2)$) is extracted and discussed. The muon spin-spin relaxation rate shows a critical slowing down (contrary to preliminary findings) with a critical exponent $n=0.7(1)$, witnessing the Ising nature of the dynamic fluctuations.  The muon precession frequencies vs. applied magnetic field reveal the saturation of the weak ferromagnetic domain structure originated by the Dzialoshinski-Moriya antisymmetric exchange.  
\end{abstract} 
\maketitle
\section{Introduction}
\label{sec:intro}

LaMnO$_3$ is a layer antiferromagnetic perovskite which displays weak ferromagnetism ($wf$). The $wf$ behaviour indicates that the ordered moments acquire a tiny ferromagnetic component by a tiny tilt out of the layer plane.
LaMnO$_3$ is also the parent of a family of magnetic materials, the manganites, which are of current wide interest because of the interconnection between their structural, transport and magnetic properties. The features of this family, including colossal magnetoresistance (CMR) and a very large spin polarization of the conduction band around 1/3 doping, make them promising for future applications. Moreover they are of prominent interest for the understanding of correlated electron systems. 

Manganites are characterized by the coexistence of several interactions on  comparable energy scales, which produce a complex interplay of magnetism, electron-lattice coupling, orbital and charge ordering. This shows up as soon as the system starts to be doped, typically by heterovalent cation substitution, giving rise to phase separation, magnetic polarons and eventually to the CMR phenomenon. 
However many of the basic ingredients, such as for instance orbital ordering and a large Jahn-Teller (JT) distortion, are already present in pure LaMnO$_3$. Some authors actually question its description as a simple insulating antiferromagnet, and suggest that the unusual layered magnetic structure is an effect of the static cooperative JT deformations on the magnetic interactions, as a sort of precursor to the double-exchange ferromagnetism of the doped compound\cite{Feinberg}.
These features underlines the relevance of an accurate determination of the properties of the end member.

Our single crystal investigation by Muon Spin Rotation ($\mu\cal SR$) attains two main results: we obtain a direct microscopic determination of a few magnetic properties of this important material and we identify with a high confidence level the interstitial sites occupied by the muons. The first aspect covers static and dynamic critical exponents of the antiferromagnetic transition, as well as the initial process of weak ferromagnetic saturation.
Critical phenomena have been addressed also in the substituted compounds, both by neutron scattering \cite{Martin,Lynn} and by macroscopic magnetization \cite{Ghosh,Mohan}, hence it is of primary interest to obtain accurate measurements of the same quantities and behaviors with a microscopic probe in the end member, much less prone to stoichiometry problem. Furthermore non trivial consequences of the weak ferromagnetism are also revealed.
The second aspect is also of general relevance, since it contributes to the understanding of muon localization in crystals, with something more than a new single case. The insight by $\mu\cal SR$ is often greatly enhanced by the knowledge of the muon site and the systematics in this field is far from complete. We believe that after our contribution the situation in LaMnO$_3$ is better understood than, for instance, in all the cuprates. Hopefully our results can be extended in part to other oxide and fluorite perovskites, including KCuF$_3$ and the high $T_c$ materials.

The paper is organized as follows: Sec. \ref{sec:exp} briefly reviews the structural and magnetic properties of the compound, some details about our sample and a brief technical introduction to $\mu\cal SR$; Sec. \ref{sec:results} illustrates our results, including the identification of the muon stopping sites based on dipolar sum calculations and on a refined fitting procedure, the zero field determination of the staggered magnetization, the dynamical critical behavior and the saturation of the weak ferromagnetic domain structure; These results are discussed in sec. \ref{sec:discussion}; Finally details on the dipolar sums and on the refined global fit are given in the appendices.

\section{Experimental}
\label{sec:exp}
\subsection{LaMnO$_3$}
\label{sec:lamno}
\begin{figure}
  \centering
  \includegraphics[width=0.4\textwidth]{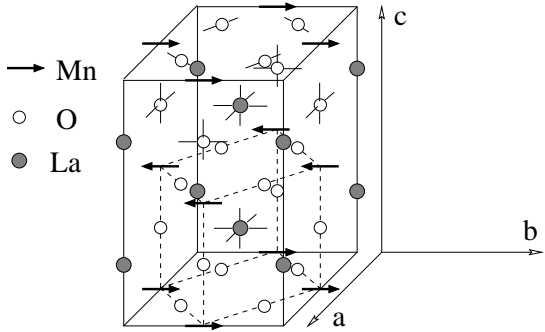}
  \caption[Orthorhombic cell of LaMnO$_3$]
          {Orthorhombic cell of LaMnO$_3$ with the A-type magnetic structure (distortions and the $wf$ canting are not shown for clarity).}
  \label{fig:cell}
\end{figure}

The crystal structure of LaMnO$_3$ is orthorhombic, belonging to space group Pnmb, with lattice parameters \cite{Huang} $a= 5.7391(2)$ \AA, $b=5.5319(2)$ \AA, $c=7.6721(2)$ \AA\ at 14 K. The unit orthorhombic cell is shown in fig. \ref{fig:cell}, together with the pseudocubic cell (dashed). In the following the subscripts $_{o,p}$ refer to these two cells, respectively. Also shown is the magnetic structure determined by neutron diffraction  below $T_N=139.5$ K. The spin is along $\hat{s}=[010]_o=[110]_p$ and the periodicity, given by $q_{AF}=(00{1\over2})_p=(001)_o$, produces a staggered  stacking of ferromagnetic planes (referred to as A-type antiferromagnetism).

The absence of the inversion symmetry around the midpoint between neighboring Mn ions along [001] implies the presence of an antisymmetric exchange, the so called Dzialoshinski-Moriya interaction\cite{Dzialoshinski,Moriya}, which produces weak ferromagnetism, i.e. a tiny canting of the spin from the orientation of a pure collinear antiferromagnet. This is indeed experimentally determined by magnetization \cite{Skumryev} and antiferromagnetic resonance \cite{AFM}, with a canting angle of 2 degrees, while neutron diffraction data\cite{Hennion} set an upper limit of 1 degree to its value. This second determination is insensitive to any accidental small fraction of the ferromagnetic compositions, hence it should reflect the intrinsic local value.  

Our sample is a microtwinned single crystal grown by the float zone method \cite{Revcolevschi} at the Laboratoire de Chimie Physique, Universit\'e Paris-Sud, cut into a slice of roughly 24x7x2 mm$^3$. Six families of cubic microtwins are present \cite{Moussa}, obtained one from another by interchange of $\pm$[100]$_p$,$\pm$[010]$_p$ and $\pm$[001]$_p$.

\subsection{The $\mu\cal SR$ experiment}
\label{sec:musr}
The experiments were performed on the GPS instrument\cite{GPS} of the Paul Scherrer Institut, Villigen (Switzerland).  The normal to the largest surface of the crystal -- close to a [111]$_o$ direction -- was oriented parallel to the muon beam direction, along the laboratory frame $\hat x$ axis. The veto counter setup (VCS) was used to minimize positron counts from muons missing the small sample\cite{GPS}. 

In a $\mu\cal SR$ experiment the positron counts from implanted $\mu^+$ decays reflect the spatial asymmetry of the decay probability, peaked around the initial $\mu$ spin direction $\hat{\zeta}=\I(0)/\abs{\I}$. Hence the count rate $N(t)=N_0\exp(-t/\tau)[1+A_0G(t)]$ is modulated by the muon polarization, i.e. the self correlation function of the muon spin projection in the counter direction $\hat \xi$, $G(t)=G_{\zeta\xi}(t)= \VEV{I_\zeta(0)I_\xi(t)}$, averaged over the implanted ensemble. Here $\tau=2.197\mu$s is the muon lifetime, $N_0$ the initial positron count rate and $A_\xi(t)=A_0 G(t)$ is often referred to as the muon experimental asymmetry function. 

For a coherent precession around a {\em transverse} magnetic field $\B\perp \hat{\zeta}$ (e.g. in a calibration $T>T_N$ experiment) one has $G(t)=\cos(2\pi\gamma \abs{\B}t)$ and one can thus measure the asymmetry parameter $A_0$ (typically around 0.26 for each GPS detector). 
A damped harmonic precession is obtained also in magnetically ordered materials, where a local field $\B$ at the muon site is produced by the electron magnetic moments. It has a dipolar contribution, $\B_{dip}$,  and a hyperfine contribution, $\B_{hf}$. The former is:
 \begin{equation}
\label{eq:dipsum}
B_{dip}^i=g\mu_B \sum_n \sum_j {{3 x^i_n x^j_n - r_n^2\delta_{ij}}\over {r_n^5}}S_n^j\end{equation} 
where $x^{i,j}_n$ with $i,j=x,y,z$ are the Cartesian coordinates and $r_n$ the modulus of the vector joining the $n$-th magnetic lattice site (with magnetic moment $g\mu_B\S_n$) to the muon site, $\rr_n=\R_n-\rr_\mu$. The hyperfine field may be written as  $\B_{hf}=\sum_j{\cal C}^j\cdot\S_j$ with $\S_j$ the spin of a nearest neighbor magnetic ion, the hyperfine path in oxides being through oxygen ions.

Muons may occupy inequivalent sites. In the ordered phase and in zero external field the presence of cubic microtwins reproduces the usual powder average:

\begin{eqnarray}    
\label{eq:polarization}
A_\xi(t)& = & \sum_i {A_0\over 3} f_i\cos\theta_\xi[2\cos(2\pi\gamma \abs{\B_i}t)e^{-t/T_{2i}}  \nonumber \\ & +  &  e^{-t/T_{1i}}]
\end{eqnarray} 
where $f_i$ is the fraction of the muon ensemble localized at a site characterized by the spontaneous local field $\B_i$; $\theta_\xi=\cos^{-1}(\hat{\zeta}\cdot\hat{\xi})$ is the angle between the initial muon spin direction and the detector axis; the transverse (precessing) component is subject to spin-spin relaxation with rate $T_2^{-1}$, while the longitudinal component is subject to spin-lattice relaxation with rate $T_1^{-1}$.

In GPS the positron detectors lie in the $\hat y$ plane (with $\hat{\xi}=\pm\hat{z}$ or $\pm\hat{x}$) and the cryostat lies along the $\hat y$ direction. A small magnetic field (5 mT) can be applied along $\hat{y}$ for calibration; larger fields may be applied along $\hat{x}$. The initial muon spin direction lies in the $\hat y$ plane at a chosen angle $170\le\psi=\cos^{-1}(\hat{\zeta}\cdot\hat{x})\le 130$ deg, depending on the settings of a cross-field device (spin rotator) in the beamline.
The spin rotator was generally {\em on} ($\psi=130$ deg) to optimize the efficiency in the VCS. 

Data analysis was performed by means of the MuZen suite \cite{muzen}.
The basic approach to data analysis was by Fourier transforming or fitting to specific models an averaged muon asymmetry function, experimentally obtained from opposite groups of detectors. E.g. with $\psi=130$ deg (spin rotator {\em on}) 
\begin{equation}
\label{eq:pol-exp}
A(t)={{N_x+N_z-\alpha(N_{\overline x}+N_{\overline z})}\over{N_x+N_z+\alpha(N_{\overline x}+N_{\overline z})}}
\end{equation}
where $\alpha$ is an average correction factor for the different detector efficiencies.
This approach loses some of the phase information and the average over a large solid angle reduces the observed muon asymmetry, but it allows the implementation of the complete set of fitting functions for $A(t)$ available in MuZen. 
A more powerful simultaneous fit of the four detectors, described in sec. \ref{sec:global}, was used for fit refinements.

\section{Results}
\label{sec:results}
Muon polarization data at three temperatures below 
$T_N$ are shown in fig. \ref{fig:p(t,T)} together with their best fit (solid curves).
Two precession frequencies are found at each temperature, indicating two magnetically inequivalent stopping sites for the muon.
\begin{figure}
  \centering
  \includegraphics[width=0.4\textwidth]{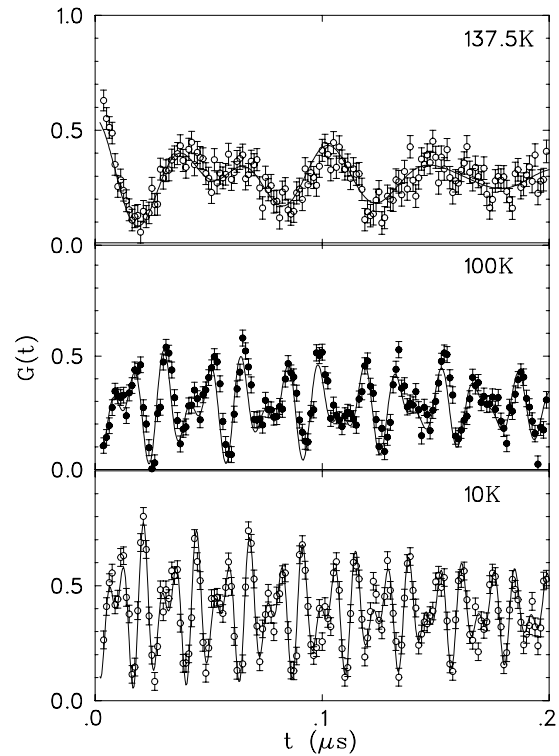}
  \caption[Muon polarization]
          {Muon polarization, as from eq. \ref{eq:pol-exp}, at three temperatures below $T_N$, displaced vertically for clarity; the solid curves are the best fits according to eq. \ref{eq:polarization}.}
  \label{fig:p(t,T)}
\end{figure}
 
\subsection{The muon sites}
\label{sec:sites}
\begin{figure}
  \centering
  \includegraphics[width=0.4\textwidth]{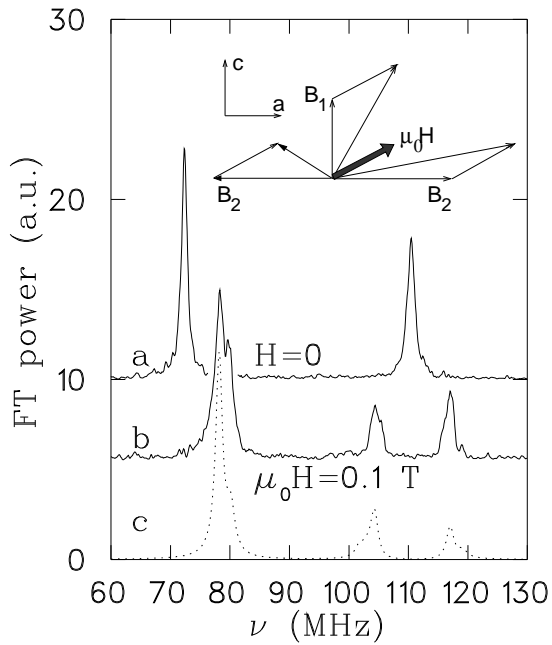}
  \caption[Fourier transform of polarization]
          {Fourier transform of polarization, simulated from the best fit parameters for different crystal orientations, identified by Euler angles $(\theta,\phi,\zeta)$ of the pseudocubic axes in the GPS reference frame. Plots are shifted both vertically and horizontally for clarity.}
  \label{fig:1kOe}
\end{figure}

The site identification is derived by comparing an analysis based on dipolar sums with the results of two experimental features 
of the internal fields probed by the muon:
their magnitude, extrapolated to zero 
temperature -- $B_1^0=0.631(1)$ T and $B_2^0=0.954(1)$ T -- and their direction. The magnitude is measured directly by the precession frequency, while the direction is revealed by the frequency splitting upon application of an external field. Since the sample is twinned there was no scope for a full scan of the angular dependence of the resulting frequencies. However the most important symmetry is already apparent from fig. \ref{fig:1kOe}, where the two top curves compare the Fourier transform (FT) of the muon polarization signal with and without an external field applied close to the [111]$_p$ direction. 
In fig. \ref{fig:1kOe} spectrum a) shows a low frequency precession around a local field $B_1=0.5339(1)$ T at 70 K, which shifts to higher frequency under an external field of 100 mT (spectrum b). At the same time the high frequency precession around $B_2=0.8150(2)$ T in spectrum a) undergoes a splitting in spectrum b). The inset shows a geometry which justifies this behaviour: the lower frequency corresponds to an internal field $B_1$ along [001] (the $wf$ axis)
while the higher frequencies correspond to an internal field perpendicular to the $wf$ axis. The external field saturates the sample and adds vectorially as shown, giving rise to one lower and two higher precession frequencies. The geometry and magnitude of the internal fields is discussed below.

\subsubsection{Low field muon site}
\label{sec:lowsite}
We start by noting that an identical behavior is detected in several $\mu\cal SR$ experiments performed on orthoferrite single crystals \cite{Holzschuh}. There the upward shift of the low frequency precession is justified on the basis of the following two assumptions: {\em a)} a local field entirely of dipolar nature, and {\em b)} a muon site within the mirror plane $z=1/4$ (or 3/4) of the orthorhombic cell. These assumptions turn out to be very reasonable for the orthoferrites, which are isostructural to LaMnO$_3$, therefore we start from the same conjecture.
Furthermore the assumption of Holzschuh {\it et al.} \cite{Holzschuh} is justified by remarking that actually point {\em a)} is a consequence of point {\em b)}. Since the muon hyperfine coupling ${\cal C}$ is transferred through oxygen and also the relevant oxygen ions belong to the mirror plane, $\B_{hf}$ vanishes by symmetry (the hyperfine contributions from two Mn ions belonging to opposite sublattices cancel).

Appendix \ref{sec:dipsum} discusses briefly the assignment: it shows that the non-vanishing dipolar field $\B_{dip}$ is along [001]$_p$ and it explains why a sufficiently large external field produces only the upward frequency shift for such a site, which is a consequence of the weak ferromagnetism of LaMnO$_3$. 

Notice that in a saturated sample, for a generic orientation of the twinned crystal, the local field at this kind of site 
\begin{equation}
\label{eq:vector}
\B_\mu=\B_{dip}+\mu_0\H
\end{equation}
should give rise to three {\em inequivalent} vector compositions (one per pair of twins sharing their $c$ axis direction), while {\em only one} upward shifted frequency is experimentally observed. The further reduction of multiplicity is due to the chosen crystal orientation which brings the three peaks in near coincidence. 

\begin{figure}
  \centering
  \includegraphics[width=0.4\textwidth]{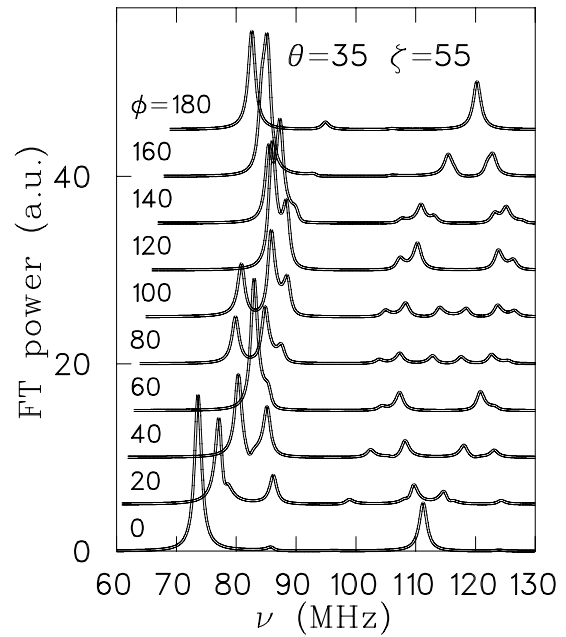}
  \caption[Fourier transform of the best fit polarization]
          {Fourier transform of the best fit polarization at $H=100$ mT, for different crystal orientations identified by Euler angles $(\theta,\phi\zeta)$ of the pseudocubic axes in the GPS reference frame. Plots are shifted both vertically and horizontally for clarity.}
  \label{fig:1kOed}
\end{figure}
By contrast fig. \ref{fig:1kOed} shows several simulations for more generic crystal orientations, where many frequency peaks are evident. Furthermore a fine structure is already apparent in all the peaks of fig. \ref{fig:1kOe}, curve b; this structure indicates that $\B$ is not exactly parallel to [111]$_p$ in this experiment and the misalignment is actually resolved by the global best fit described in sec. \ref{sec:global}.

In order to corroborate our assumptions for the low field site we performed a numerical evaluation of the dipolar sums (eq. \ref{eq:dipsum}), considering the low temperature crystal structure \cite{Huang,Moussa} and a moment of 4$\mu_B$ per Mn ion \cite{moment}. We evaluated the sums over a large spherical domain centered at the muon site, with a numerical error of less than 0.5 mT.

\begin{figure}
  \centering
  \includegraphics[width=0.4\textwidth]{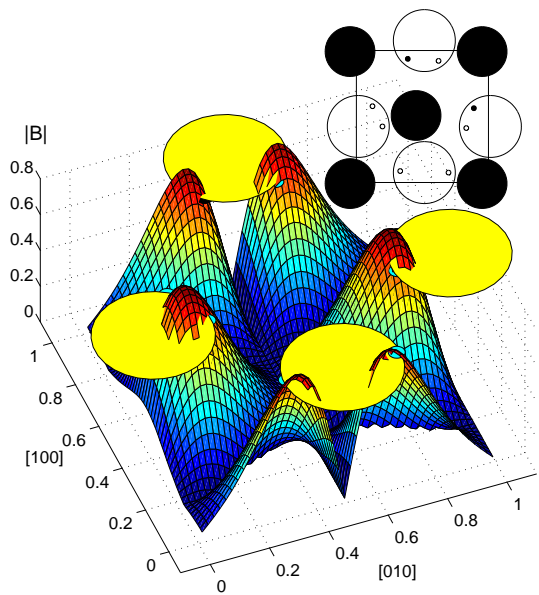}
  \caption[Field surface in the mirror plane]
          {$\abs{\B_{dip}}$ as a function of the muon position in [001]$_p$ mirror plane at $z=1/4$; The large shaded circles of radius 1.1\AA\ are centered at O sites; The inset shows the mirror plane with O (open circles), La (filled circles) and candidate mu sites (open symbols, $\cal H$ sites are filled).}
  \label{fig:plotfields}
\end{figure}
Fig. \ref{fig:plotfields} shows the magnitude of $\abs{\B_{dip}(x,y)}$ as a function of the muon position $(x,y)$ in the (001)$_o$ plane at $z=1/4$; the large solid circles correspond to a radius of 1.1 \AA\ centered on the O ions, equal to the average $\mu$-O distance for oxides \cite{Schenck}, hence the muon site is expected to lie on its boundary. Candidate sites lie at the intersection of these curves with the field surface at the experimental value of $0.631(1)$ T.
The inset shows their position in the mirror plane.
 
One site, [0.389 0.937 0.25], is marked by a darker symbol, together with its symmetry replica, [0.889 0.563 0.25]. It very closely corresponds to the main site identified in the orthoferrites (see fig. 10, site 2, in ref. \cite{Holzschuh}): they are farthest away from positive La ions, facing a large empty space and their cell coordinates also roughly agree. The correspondence implies that the internal field at this muon site scales with both lattice parameters and transition metal magnetic moments as dictated by the dipolar coupling. Hence the combined orthoferrite and LaMnO$_3$ experiments corroborate that the hyperfine contributions must be negligible.
We shall refer to these and their replica in the $z=3/4$ mirror plane as the Holzschuh ($\cal H$) site.

Holzschuh {\em et al.} calculated that also along the [{$1\over2$} 0 z)$_p$ line $\B_{dip}$ has the wanted field orientation. However along this locus a consistent hyperfine contribution is required to agree with the experimental internal field value. Since the hyperfine term in general is not collinear to the dipolar term, the required experimental symmetry would be lost. The $\cal H$ site alone can satisfy both magnitude and direction constraints.

The exact vector composition of $\mu_0\H$ and $\B_{dip}$ for the experiment of fig. \ref{fig:1kOe} were obtained from a global fit of the $\mu\cal SR$ data -- described in detail in appendix \ref{sec:global} -- which yields the following $\mu\cal SR$ assessment of the crystal orientation: the three pseudocubic axes had director cosines of [0.74(3) 0.51(3) 0.43(4)] with the external field. 
This is  in reasonable agreement with a subsequent neutron diffraction determination [0.75(6) 0.57(6) 0.36(6)], where the large error bars come mainly from systematic errors of $\sim 4$ deg in the sample realignment between the two experiments. 

\subsubsection{High field muon site}
\label{sec:highsite}
Fig. \ref{fig:1kOe}b reveals that the application of the external field in the chosen experimental geometry gives rise to two precessions equally separated on either sides of the zero field high frequency peak. Thus {\em i)} the high frequency site does not belong to the mirror plane and {\em ii)} only two main peaks are observed. This implies again a highly symmetric site since 
the {\em generic} site in a Pnmb cell (type $d$ in Wyckoff notation\cite{Inttables}, with multiplicity 8) in a {\em generic} orientation of the twinned crystal in the external field would give rise, according to eq. \ref{eq:vector}, to 48 inequivalent vector compositions! Their number reduces greatly thanks to the chosen crystal orientation, but the experimental results are compatible only with the net internal field $\B_{dip}+\B_{hf}$ along one of the pseudocubic axes. This is shown in fig. \ref{fig:1kOed} where Fourier spectra of the best fit function are simulated for different orientations of the crystal in the external field. It is clear that 
both the high symmetry of the muon site and the best-fit crystal orientation are rather stringent conditions: if the two were not simultaneously met the spectrum would be spread out into many subpeaks, mostly below noise level. We therefore conclude that the internal field $\B_{dip}+\B_{hf}$ lies along the local [110]$_p$=[100]$_o$ direction, i.e. the direction of the ordered moments.

\begin{figure}
  \centering
  \includegraphics[width=0.4\textwidth]{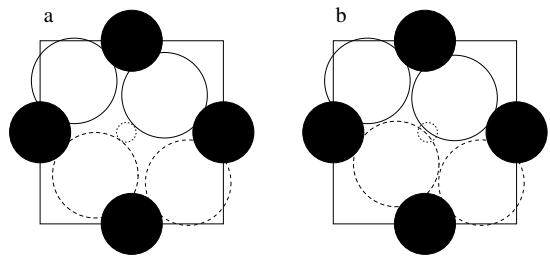}
  \caption[High site]
          {The basal plane a) in the undistorted cell; b) when an O-Mu-O center is formed in $\cal C$. Mn ions (filled circles); O ions (open circles) are centered 0.03 \AA\ above -- solid -- or below -- dashed -- the plane; the $\cal C$ site (dotted circle); the points on the solid squares are also loci where $\B_{dip}+\B_{hf}$ is parallel to [100]$_o$ (see text).}
  \label{fig:highsite}
\end{figure}

 Assuming isotropic {\em hf} coupling, both $\B_{dip}$ and $\B_{hf}$ must be parallel to [100]$_o$. This particular field orientation is obtained along the following lines in the Pnmb cell: [x 0 0]$_o$, [0 y 0]$_o$, [x 0.5 0]$_o$, [0.5 y 0]$_o$, [0.5 0 z]$_o$, and [0.5 0.5 z]$_o$, plus those equivalent by symmetry. Most of them coincide with the oxygen-cation bonds, thus being unlikely locations for a positive muon. However the [x 0.5 0]$_o$ and [0.5 y 0]$_o$ lines connect two magnetic ions and intersect at the center of the basal plane, in $\cal C$=[0.5 0.5 0]. The region around $\cal C$ is empty and it is a possible location for the high frequency muon site. Site $\cal C$ is shown in fig. \ref{fig:highsite}a, together with a few of the mentioned lines.

There is an additional reason to consider site $\cal C$ in particular. It is well known that muons implanted in several fluorides\cite{Brewer} end up in a F-Mu-F structure, where two F ions are drawn towards the muon sitting at an interstice midway between them, in a muon version of the hydrogen bond. The bond distance in this center is remarkably the same (1.16 \AA) throughout a large series of compounds. We have recently observed\cite{kcufff} the same center in KCuF$_3$, which is essentially isostructural to LaMnO$_3$. We suggest that an analogous O-Mu-O center could be formed here, exactly at site $\cal C$, as shown in fig. \ref{fig:highsite}b. The resulting overlap with O could easily justify the required hyperfine contribution of $B_{hf}=0.51$ T, in addition to the calculated $B_{dip}(C)=0.41$ T. 

Finally one may note from the inset of fig. \ref{fig:plotfields} that the O ions in the mirror planes cannot be drawn together because of the La ion hindrance, whence the $\cal H$ site remains locally favored in that plane.

\subsection{Critical behaviour}
\label{sec:magne-relax}
\begin{figure}
  \centering
  \includegraphics[width=0.4\textwidth]{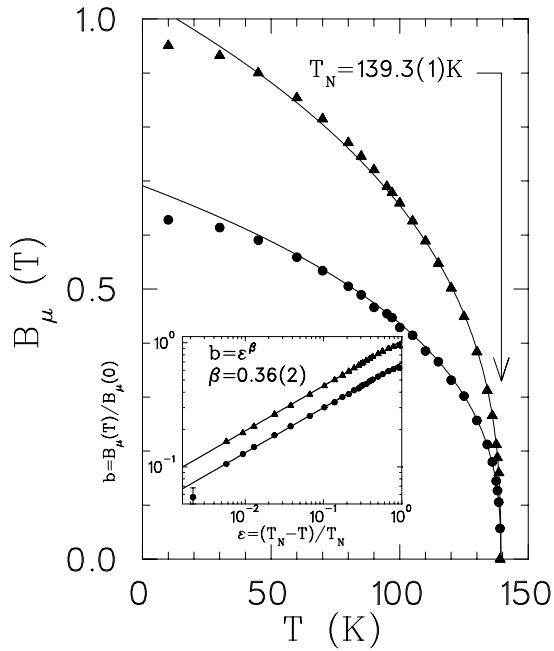}
  \caption[B(T)]
          {Temperature dependence of the $\cal H$ and $\cal C$ internal fields. The solid lines are the best fit of the critical behavior. The inset replots the data on a log-log scale; the solid lines are the same best fits.}
  \label{fig:B(T)}
\end{figure}

The zero field experiments provide a measurement of the temperature dependence of the staggered magnetization. Figure \ref{fig:B(T)} shows the behavior of the internal fields at both muon sites. 
They are proportional to the local magnetic moment on Mn, i.e. to the magnetic order parameter, by means of the coupling to the muon spin, as shown e.g. in eq. \ref{eq:dipsum}. Hence we may extract the critical exponent of the order parameter, $\beta$, from them; to this end they are replotted as a function of reduced temperature $\varepsilon=1-T/T_N$ on a log-log scale in the inset of fig. \ref{fig:B(T)}. 
The data follow a power law $\varepsilon^\beta$ on a remarkably large temperature interval, at least up to $\varepsilon=0.6$. The critical exponent extracted from this interval, $\beta=0.36(2)$ is that predicted by the 3D Heisenberg model \cite{Frey}, within experimental errors.  

Incidentally, this analysis reveals a perfect agreement  between the two muon sites, which confirms that their occupancy is stable in the observed temperature interval.

We now turn to the muon relaxations which give access to the dynamical aspects of the critical behaviour  \cite{derenzia,derenzib,brown}. Approaching $T_N$ from below fluctuations the AF order appear. The closer one gets to $T_N$, the larger are the fluctuating clusters and the slower is the rate of spin fluctuation, both effects contributing to a diverging spin relaxation.

In preliminary reports \cite{AMR,musr99} we wrongly remarked the absence of a critical divergence of the 
relaxation rates in LaMnO$_3$. We had overlooked that in those early data the sample had experienced an external field at low temperature, resulting in inhomogeneous broadening effects which masked the critical divergence.
\begin{figure}
  \centering
  \includegraphics[width=0.4\textwidth]{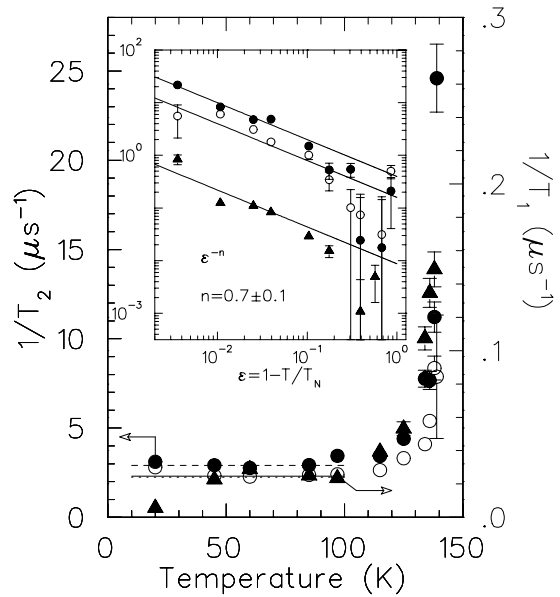}
  \caption[Critical divergence of the relaxation rates]
          {Critical divergence of the transverse and longitudinal relaxation rates; circles: $T_2^{-1}$ (open -- $\cal H$ site, filled -- $\cal C$ site), solid triangle $T_1^{-1}$. The inset shows the critical part on a log-log plot.}
  \label{fig:decays}
\end{figure}
Figure \ref{fig:decays} shows instead that in proper zero field experiments 
both $T_1^{-1}$ and $T_2^{-1}$ diverge at $T_N$ as $\varepsilon^{-n}$.  
By subtracting the non-critical contribution (indicated by the dashed lines in the figure) a dynamical critical exponent $n=0.7(1)$ may  be extracted from the relaxation rates\cite{Heller,derenzia,derenzib}, which is in agreement with the theoretical best estimates \cite{Frey} based on the Ising model ($n\approx2/3$ -- cfr. with Heisenberg model, $n\approx0.35$). This is indeed what is expected, since the anisotropy of the interaction dominates in the critical regime\cite{Bucci}.

\subsection{Weak ferromagnetic domain structure}
\label{sec:weakfm}

\begin{figure}
  \centering
  \includegraphics[width=0.4\textwidth]{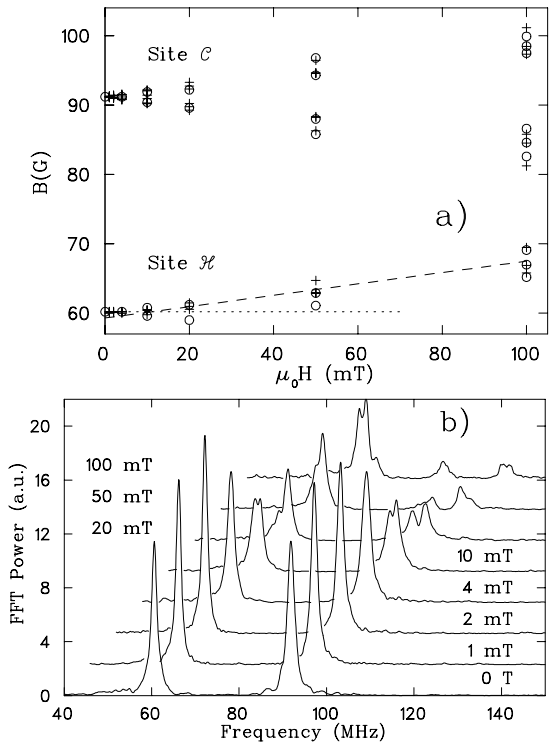}
  \caption[B(H)]
          {Longitudinal field data: a) Frequencies obtained from the global fit (crosses) and from the FT (open circles); the lines are guides to highlight the low (dotted) and high (dashed) field behaviour; b) FT spectra of the data (displaced diagonally for clarity).}
  \label{fig:B(H)}
\end{figure}

Fig. \ref{fig:B(H)}b shows the FT spectra of several data sets recorded in increasing external longitudinal fields ($1 \le \mu_0 H\le 100$ mT, apodization exponential filter of 100 kHz) and fig.\ref{fig:B(H)}a displays the field dependence of the corresponding frequencies. 

Above 20 mT the features described in details in sec. \ref{sec:sites} are apparent from both plots: the frequency belonging to the $\cal H$ site shifts upwards and eventually gives rise to three satellites, while the $\cal C$ site gives rise to two groups of satellites on either sides of its zero field resonance.

The extrapolation of this high field behaviour to low fields evidences an effective saturation induction, $\mu_0 H_s=7\pm 1$ mT. This is best seen from the average field dependence of the $\cal H$ peaks (although it holds also for the $\cal C$ peaks). The dashed line is optimized on the high field values of the central fit frequency, while the low field data agree better with the constant dotted line. The quoted demagnetization is taken from the intercept of the two lines. 
This behaviour resembles that of a soft magnet, although LaMnO$_3$ is definitely not soft: the total magnetic energy, $\propto \B\cdot\H$ is minimized for $B=0$, hence external fields $H_0<H_s$ are shielded inside the sample.

\section{Discussion and conclusions}
\label{sec:discussion}
We have definitively confirmed the $\cal H$ site location for pseudocubic perovskites first suggested by Holzschuh {\em et al.} \cite{Holzschuh} in orthoferrites. This indicated that the analogy between manganites and orthoferrites is worth pursuing. 
Holzschuh {\em et al.} detected three magnetically inequivalent sites for the muon. The missing fraction in our experiment probably justifies a third site, with either a very large internal field (a conservative estimate of the detection frequency-passband cut-off is $400$ MHz) or very large relaxation rates. Holzschuh {\em et al.} discuss the metastability of their high frequency sites, which depopulate above $\approx 150$ K in favor of the low frequency $\cal H$ site. We cannot directly confirm this feature since $T_N$ in LaMnO$_3$ is too low and the two observed sites are both stable up to that temperature.

We have found a strong analogy between the high field site and the F-Mu-F center detected in fluorides \cite{Brewer}. As a matter of fact the formation of an O-Mu-O center, the muon version of the hydrogen-bonding, is a very natural expectation for partially ionic crystals.

After the first draft of the present paper we became aware of a preprint \cite{HeffnerB} on $\mu\cal SR$ in La$_{1-x}$Ca$_x$MnO$_3$
powders with $0\le x\le 1$. They confirm our earlier reports of two sites\cite{AMR,musr99} in LaMnO$_3$ and agree on the $\cal H$ site attribution. Their data, however, are consistent with two sites only, while they observe three sites in the other end-member, CaMnO$_3$.   

Let us now discuss briefly the critical behaviour as seen from the local fields and from the relaxations. The log-log plots of fig. \ref{fig:B(T)} (inset) shows a remarkably large range on which the $\varepsilon^\beta$ behaviour holds, and the best fit for both sets yields with $T_N=139.3(1)$ K and $\beta=0.36(2)$.
A cross-over from Heisenberg ($\beta=0.365(3)$) to 3D-Ising behaviour\cite{Frey}  ($\beta=0.325(2)$), is expected close enough to $T_N$, since the anisotropy which is present in the ordered phase must eventually dominate, for very small $\varepsilon$. This cross-over is not directly observed in our magnetization data, but this is hardly significative, in view of the small variation of $\beta$ between the two limits. 
The Ising-type behaviour does indeed show up in the relaxation rates of fig. \ref{fig:decays}, whose critical exponent (related \cite{derenzib} to the dynamical exponent $z$) displays a value, $n=0.7(1)$, in full agreement with 3D-Ising predictions. 

We must mention that the muon determination of the temperature dependence of the staggered magnetization, $M_\mu(T)$, does not fully agree with that obtained by neutron diffraction, $M_n(T)$, on the same crystal \cite{Moussa}.
Assuming that the two data sets agree at $T=0$, where the muon site attribution is fully consistent with $\approx 4 \mu_B$ on Mn determined by neutrons, the deviation -- $M_\mu<M_n$ -- increases with temperature. 
Although the origin of this discrepancy is not understood,
we assume that both techniques measure intrinsic sample properties. We hope that our planned Mn NMR experiments will help clarifying this issue.

Let us finally discuss the observed $wf$ saturation. For an ideal soft wire one would have $H_s=M_s$. Since the sample is twinned we must expect a shielding field from the macroscopic average over a large number of cubic twins to mimic that of a soft material, albeit with saturation magnetization $M_s/3$. This reduction factor is most clearly explained thinking of a twinned single crystal with $\H_0\parallel [001]$, the easy $wf$ axis for one family of twins: only their contribution, $1/3$ of the total, produce a shielding.
Since LaMnO$_3$ has a large anisotropy, only the projection of the external field $\H_0$ along the easy axes may be shielded by demagnetization, hence $H_s\cos\theta=M_s/3$, where $\theta$  is the angle between the easy $wf$ axis and $\H$ (in our experiment the average of $\cos\theta$ is 0.53). 
 Approximating the sample shape to an ellipsoid, with demagnetizing factor $N$, one finally has that $H_s=N M_s/(3\cos \theta)$. An estimated value of $N=0.17$ may be derived from standard formulas\cite{Becker}, to yield $M_s=3.4(6)$ kA/m, i.e. 0.22(4) $\mu_B$ at 90 K, consistent with magnetization measurements \cite{Skumryev}. This implies a canting angle of order 2 degrees, larger than the 1 deg upper limit set by neutrons\cite{Hennion}. However our determination is essentially equivalent to a magnetization measurement and it is not a true microscopic determination, although it is detected via a local probe.

A closer inspection at the 10 and 20 mT spectra of fig. \ref{fig:B(H)} is in qualitative agreement with this picture, with a small further coercive effect:
at 10 mT an incipient splitting of the $\cal H$ frequency is apparent, with a negligible domain polarization, while at 20 mT a low frequency shoulder of the $\cal H$ peak still appears, from the minority domains.

In conclusion we have clarified the identification of the two muon sites in LaMnO$_3$, the $\cal H$ and $\cal C$ sites. Our single crystal data provide a definitive assessment of the former, while the latter is proposed here for the first time. The $\cal C$ site, analogous to what is observed in many fluorides, might be more frequent than previously thought. These assignments allow us to extract an estimate of the $wf$ canting angle. The critical behaviour of this material agrees with its current view as a Heisenberg system with a cross-over to Ising behaviour due to anisotropy. 

\appendix

\section{Dipolar sums}
\label{sec:dipsum}
Let us consider the dipolar contribution to local field at a muon site, given by eq. \ref{eq:dipsum}.
In the ordered state one may replace all $S_n^j$ with 
either $S_\alpha^j$ or $S_\beta^j$, depending on whether $n$ belongs to sublattice $\alpha$ or $\beta$, to yield
\begin{equation}
\label{eq:diptensor}
B^i_{dip}= D_\alpha^{ij}S_\alpha^j+D_\beta^{i,j}S_\beta^j
\end{equation}
where the two tensors $\hat{D}$ represent the sum of eq. \ref{eq:dipsum} restricted to each sublattice. For a collinear antiferromagnet (a very good approximation for LaMnO$_3$ in view of the minute canting) $\S_\beta=-\S_\alpha=\S$, hence $\B=(\hat{D}_\alpha-\hat{D}_\beta)\cdot\S$ and one may refer simply to $\hat{D}=\hat{D}_\alpha-\hat{D}_\beta$. 

\begin{figure}
  \centering
  \includegraphics[width=0.4\textwidth]{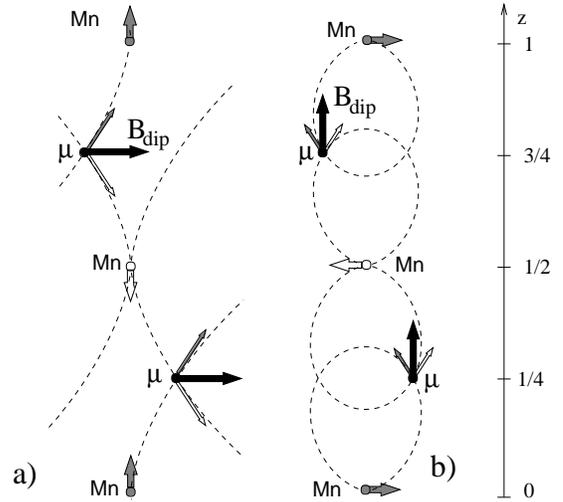}
  \caption[Symmetries of the local dipolar tensor]
          {Symmetries of the local dipolar tensor at equivalent muon sites in the mirror planes $z=1/4,3/4$; three Mn moments along $\hat c$ are shown (open and grey circles with thick spin arrows) with their dipole field lines (dashed) through the muon sites. Other arrows represent the local magnetic field at the muon from a pair of symmetric moments: a) Mn spin component along $\hat{c}$; b) Mn spin component normal to  $\hat{c}$, in the plane containing the muon sites; The third component (not shown) gives vanishing contribution by symmetry.}
  \label{fig:tensor}
\end{figure}
Let us consider first the low frequency site. Figure \ref{fig:tensor} graphically shows that the total dipolar tensor $\hat{D}$ for muon sites in the mirror plane $z=1/4,3/4$ of the Pnmb cell possesses the following symmetry:
\begin{equation}
\label{eq:tensor}
\hat{D} =
\left[ 
\begin{array}{ccc}
0 & 0 & d_3 \\ 
0 & 0 & d_2 \\ 
d_3 & d_2 & 0
\end{array}
\right]
\end{equation}
which guarantees that, for $\hat{s}=[110]_p$, $\B_{dip}$ lies along $[001]_p$. 

If we neglect the tiny spin canting the Mn moments lie in the [001] plane, hence  we can concentrate on  fig. \ref{fig:tensor}b), which illustrates the relevant case for both mirror planes. 
Since the two Mn sublattices are symmetric with respect to the mirror plane, the dipolar field properties are identified considering just two equivalent Mn ions, one per sublattice. Only the components of the Mn magnetic moments in the plane of the drawing are considered, since the perpendicular components give opposite dipolar field contributions which cancel.
The thinner arrows in the figure indicate the dipolar term from each Mn ion at a generic site in the mirror plane: it is graphically evident that their sum lies along [001], as predicted by eq. \ref{eq:tensor}. Similar arguments hold for the other components of the tensor.


Notice that the figure shows just one AF domain (the bottom Mn moment belongs, say, to the $\alpha$ sublattice); the opposite direction of $B_{dip}$ is obtained in the other AF domain (where the bottom Mn moment belongs to the $\beta$ sublattice; imagine the figure with all arrows reversed). However according to the Pnmb symmetry the $\beta$ domain must also have its $wf$ component in the opposite direction along $[001]_p$.

Based upon these consideration we may understand the field behavior of the $\mu\cal SR$ precessions. The $wf$ spin component is totally negligible in our dipolar sums, but it gives a net $wf$ moment $\M^{wf}$ to each domain, which couples to an external field $\H$. When $\H$ saturates the macroscopic $wf$ moment only one of the two domains survives -- specifically that with $\H \cdot \M^{wf}>0$. It appears that $M^{wf}$ and $B_{dip}$ at the $\cal H$ muon site are pointing in the same direction, so that
when the component of $\H$ along [001] is larger than the saturation value one has $\H\cdot\B_{dip}>0$, hence the precession frequency shifts upwards, as in fig. \ref{fig:1kOe}. 
By contrast at the $\cal C$ site the internal field is in the direction of the spin and its the vector composition with $\mu_0\H$ is not influenced by the AF domain structure (all possible geometries are obtained within each domain).

\section{Global fit}
\label{sec:global}
We have employed a procedure to fit the whole set of data recorded at each temperature and field. The global chi square, 
$\chi^2=\sum_{i,\alpha}(N^\alpha(t_i)-N_{th}^\alpha(t_i))^2/N^\alpha(t_i)$, is
obtained from the count rates $N^\alpha$ of the four detectors directed along $\alpha=\pm\hat{x}, \pm\hat{y}$, at each recorded time $t_i$. The model function $N_{th}$, briefly described in sec. \ref{sec:musr}, depends on several parameters, of which only few -- field intensities, their orientations and the relaxation rates -- are informative of the magnetic properties of LaMnO$_3$. The rest are of lesser interest in this respect, such as the angle $\psi=\cos^{-1}(\hat{\zeta}\cdot\hat{z})$ defining the initial muon spin direction, the experimental asymmetry $A$ of the reference $\hat x$ detector, the muon fractions, $f_i$ and the individual initial count rates, $N^\alpha_0$, instrumental delay times, $t^\alpha$ and relative efficiencies $e^\alpha$ for all the $\alpha$ detectors. The asymmetry parameter of each detector is $Af_ie^\alpha$, and the efficiency of the reference detector is fixed, $e_x\equiv 1$. All these are a byproduct of the fit procedure, and a number of them may be accurately determined in separate high statistics calibration experiments.

All models generally include one or more contribution to the polarization functions, as in the case described by eq. \ref{eq:polarization}, which points out that quite generally each local field $\B_i$ identifies a muon fraction $f_i$ 
characterized by its transverse and longitudinal relaxation rates, $1/T_{2i}$ and $1/T_{1i}$. When fitting more than one fraction, however, it is often impossible to distinguish individual {\em longitudinal} rates and a single $1/T_{1}$ value is employed.

The advantage of the global approach is that the direction of the local field $\B_i$ may be extracted from the data, since it determines the aperture of the muon spin precession cone. This is incidentally not true for the zero-field sum over our cubic twins, discussed in eq. \ref{eq:polarization}, since the information on the local field direction is lost by symmetry. Nonetheless this zero field model is very easy to implement in the global fit scheme and the field values of section \ref{sec:magne-relax} are indeed obtained from such a fit.
 
The global approach becomes more valuable in an applied external field $\H$, were the direction of the internal field $\B_{hf}$ may be determined. In this case the direction of the local field $\B_{hf}+\mu_0\H$ in each twin may be expressed in terms of the three Euler angles $(\theta,\phi,\zeta)$ which define the crystal orientation in the laboratory $\hat x \hat y \hat z$ frame.

Therefore, following the discussion of appendix \ref{sec:dipsum}, we have implemented two further models: one to account for the $\cal H$ site in the saturated $wf$ state (spontaneous field $\B_{dip}$ along the local $c$ direction, in three twin replicae, chosen with the condition that $\B_{dip}\cdot\H>0$); another one to account for the $\cal C$ site (spontaneous field $\B_{hf}$ along the local electron moment direction -- $[010]_o$ -- in six distinguishable replicae). 
The analytic expression of the corresponding functions $N^\alpha_{th}(t)$, involving their dependence on $(\theta,\phi,\zeta)$ and the vector composition of the internal and external fields, was generated and checked by means of a simple Maple V program. 

The best-global-fit values discussed in section \ref{sec:sites} refer to a model function including both sites. Because of the relatively large number of parameters the convergence of the global fit requires 
some practice, but reliable fits were always obtained with $\chi^2$ values better than 1.1 per degree of freedom. 

The fit quality is good but not perfect, as it may be judged from the comparison between the FT of the data and of the fitting function in fig. \ref{fig:1kOe}b,c. The small disagreement may be imputed to simplifying assumptions in the fit function, in which the crystal twins are weighted equally, while x-ray diffraction in similar samples showed slight preponderance of one twin \cite{Skumryev}, and relaxations are all assumed Lorentzian, while probably a Gaussian $T_2$ rate is more appropriate. 

The behaviour of some of the less interesting fit parameters confirms the soundness of the model. 
\begin{figure}
  \centering
  \includegraphics[width=0.4\textwidth]{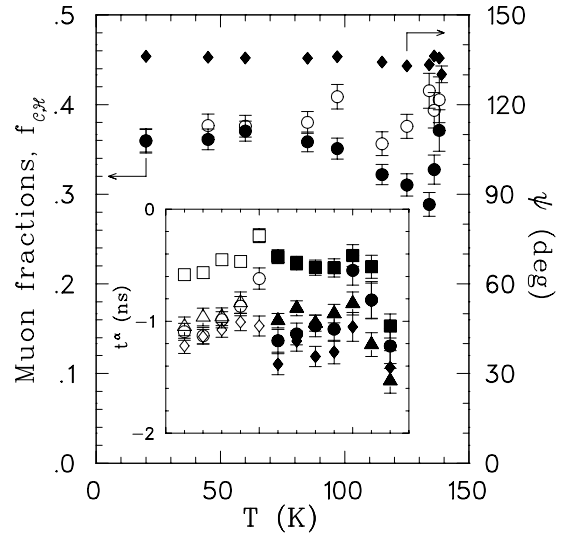}
  \caption[Parameters of the zero field global fit]
          {Some parameters of the zero field global fit: muon fractions at the two sites and initial muon phase. The inset compares the detector time delays in  zero field (open symbols) and in longitudinal fields (filled symbols).}
  \label{fig:angles}
\end{figure}
Figure \ref{fig:angles} shows the temperature dependence of the muon  fractions and of the initial muon spin phase $\psi$, obtained from the zero field data. The phase $\psi$ is perfectly stable, while the two fractions do show some small spurious correlation effects close to the transition, where the relaxation rates diverge. This is acceptable since it has no appreciable influence on the critical analysis of sec. \ref{sec:magne-relax}. The inset shows the scattering of the time delays of the four detectors both in the zero field data (open symbols) and in the longitudinal field data, which are fitted with different functions. These parameters are stable and  
insensitive to both field and model function employed.


\begin{thebibliography}{99}
\bibitem{moment} The $T=0$ magnetic moment is measured as 3.7$\mu_B$ and 
3.9$\mu_B$ from ref. \cite{Huang,Moussa} respectively, with a certain sample dependent difference, but a spin-only $S=2$ state is expected for the ideal crystal. We assumed the full theoretical value.
\end{thebibliography}
\end{document}